\documentclass[preprint]{aastex}
\usepackage{apjfonts}
\usepackage{graphicx}
\usepackage{epsfig}
\usepackage{float}
\usepackage{amsmath}
\usepackage{natbib}
\usepackage{float}
\usepackage{amssymb}

\usepackage[usenames]{color}

\shorttitle{}
\shortauthors{}

\begin{document}
\title{A state-dependent influence of the type-I bursts on the accretion in 4U 1608--52?}

\author{
Long Ji\altaffilmark{1}, Shu Zhang\altaffilmark{1}, YuPeng
Chen\altaffilmark{1}, Shuang-Nan Zhang\altaffilmark{1}, Diego F. Torres\altaffilmark{2,3}, Peter
Kretschmar\altaffilmark{4}, Jian Li\altaffilmark{1}}
\altaffiltext{1}{Laboratory for Particle Astrophysics, Institute of High Energy
Physics, Beijing 100049, China}
\altaffiltext{2}{Instituci\'o Catalana de Recerca i Estudis Avan\c cats (ICREA),
08010 Barcelona, Spain}
\altaffiltext{3}{Institute of Space Sciences (IEEC-CSIC), Campus UAB, Torre C5,
2a planta, 08193 Barcelona, Spain}
\altaffiltext{4}{European Space Astronomy Centre (ESA/ESAC), Science Operations
Department, Villanueva de la Ca\~nada (Madrid), Spain
}

\begin{abstract}
We investigated the possible feedback of type-I burst to the accretion process during the spectral evolution of the atoll source 4U 1608--52.
By fitting the burst spectrum with a blackbody and an adjustable, persistent spectral component, we found that the latter is significant state-dependent.
In the banana state the persistent flux increases along the burst evolution, while in the island state this trend holds only when the bursts are less luminous and starts to reverse at higher burst luminosities.
We speculate that, by taking into account both the Poynting-Robertson drag and radiation pressure, these phenomena may arise from the interactions between the radiation field of the type-I burst and the inner region of the accretion disc.

\end{abstract}

\section{Introduction}
Type-I bursts are nuclear explosions triggered by unstable burning on the surface of neutron stars
\citep[for details see e.g.][and references therein]{Lewin1993,Galloway2008}. The burst spectrum is usually morphologically described as a pure blackbody,
assuming that the persistent emission remains unchanged and can be subtracted off during bursting \citep[see e.g.][]{vanParadijs1986,Kuulkers2003}.
Generally, the persistent emission for a neutron star X-ray binary (XRB) mainly consist of the spectral components of a multicolor blackbody, a blackbody, and a powerlaw that is believed to have origins of accretion disc, boundary layer (between the accretion disc and the neutron star surface),  corona or  jet \citep[see e.g.][and references therein]{Belloni2010, Darias2014}.
In this letter, we investigated a well-known XRB 4U 1608--52. According to the spectral and timing properties, 4U 1608--52 was classified as an atoll source \citep{Hasinger1989}, for which the typical outburst evolution experiences the so-called  island and banana states.

Theoretical predications for the possible deviations of the burst spectrum from a blackbody have been made under different mechanisms.
For instance, the reflection of the burst emission by the accretion disc was considered in \citet{Ballantyne2004}.
In addition, the radiation of the boundary layer may also be affected by type-I bursts \citep{intZand2013}.
\citet{Walker1992} proposed that the radiation torque may induce a substantial increase of the accretion rate during a type-I X-ray burst, likely resulting from the effect of Poynting-Robertson drag.
On the contrary, \citet{Kluzniak2013} predicted an ejection under the radiation pressure, of a large part of optically thin gas orbiting the neutron star.

In observations, the possible influence of the bursts upon the accretion disc is mostly investigated with the so-called photospheric radius expansion (PRE) bursts.
For PRE bursts, the X-ray luminosity reaches local Eddington limit and is strong enough to lift up the outer layers of the neutron star \citep{Galloway2008}.
\citet{intZand2013} detected a PRE burst from SAX J1808.4-3658 in  an observation campaign of {\it Chandra} and {\it Rossi X-Ray Timing Explorer} ({\it RXTE}), and found spectral deviations from a blackbody at both low and high energies.
In addition, by assuming that the shape of the persistent spectrum remains stable throughout the burst and involving a dimensionless quantity "$f_{\rm a}$" to account for the change in normalization of the persistent emission (hereafter $f_a$ model), \citet{Worpel2013} analyzed 332 PRE bursts born out of 40 atoll sources. They found that the majority of the best-fit values of $f_{a}$ are significantly greater than 1, suggesting an enhancement of the persistent flux during these bursts.
Since the changes of the structure of the accretion disc, i.e. the transitions between the standard disc and the advection-dominated accretion flow (ADAF), are expected to influence the burst spectrum, the variability of the persistent flux observed in PRE burst may have a dependence on the spectral states.
However, most of the PRE bursts are located in the banana states \citep{Watts2012} and the influence of the island non-PRE bursts on accretion process is less known. In order to study the bursts' influence thoroughly, one needs to account as well for the non-PRE bursts occurring mostly in the island state.
To this end we studied the well-known X-ray binary 4U 1608--52, which was firstly detected by {\it Vela-5} and subsequently observed with {\it Uhuru}, {\it HEAO-1}, {\it RXTE} and other satellites \citep{Belian1976,Tananbaum1976,Fabbiano1978,Galloway2008}.
The distance to 4U 1608--52 was estimated as $\sim$ 3.2 kpc with PRE bursts when their luminosities reach Eddington limit, under the assumptions of   a canonical neutron star with $M = 1.4 M_{sun}$, $R=10$ km, and $X=0.7$ \citep{Galloway2008}.


\section{Observations and data analysis}
We analyzed all {\it RXTE} Proportional Counter Array \citep[PCA,][]{Jahoda1996} observations of 4U 1608--52 and found 46 bursts, 19 of which exhibited PRE. The PRE bursts have been identified by \citet{Poutanen2014}.
The Standard 2 mode data were used when studying the persistent spectra, for their better calibration, and the Event mode data were adopted when studying detailed properties of bursts, because of their high resolution.
There are five co-aligned Xe multiwire proportional counter units (PCUs) aboard {\it RXTE }, and all active ones were used for the following analysis.

The spectral fits were performed using {\it XSPEC} version 12.7.1, during which a multiplicative model component (wabs in XSPEC) was employed to correct for the effects of interstellar absorption and the hydrogen column density was fixed to $0.89 \times 10^{22}$ atoms cm$^{-2}$ \citep{Keek2008}.
The instrumental background of PCA was estimated with the latest bright source models and generated using the {\it pcabackest} version 3.8, {\it heasoft} release 6.12.
In the spectral fits, the Levenberg-Marquardt algorithm was used to find the local best-fit values and their corresponding confidence intervals.
Because the effective area of PCA drops off rapidly above 20 keV, sometimes only a few counts are observed in higher energy band.
To ensure that we could get an unbiased estimate of the parameters, we have used the likelihood method, namely, the command "statistic cstat" in XSPEC.
The fitted energy range was in principle 3--25 keV but subjected to a modification to 3--20 keV or 3--15 keV to avoid unphysical results born out of the poor statistics.

The spectral states in which the type-I bursts occur were inferred from the positions in the color-color diagram (CCD). In practice, we first produced PCA lightcurves for each energy band (2.2--3.6 keV, 3.6--5.0 keV, 5.0--8.6 keV and 8.6--18.8 keV) in a 128 s time bin, during which we excluded the data affected by type-I bursts.
To correct for the PCA gain drift over time because of the changing voltage settings, especially for different {\it RXTE} gain epochs, we used the Crab to calibrate the channel-to-energy conversions of PCA by assuming that its intensity remains constant \citep{Belloni2010}. The small-scale variation of the Crab \citep{Wilson2011} has little effect on the following analysis.
Similar to \citet{Galloway2008}, we took the count-rate ratios of  3.6--5.0 keV to  2.2--3.6 keV  as soft color and 8.6--18.8 keV  to  5.0--8.6 keV  as hard color for the input of the CCD diagram (see figure 1).
Finally,  the "island state" or the "banana state" was defined with a  hard color smaller or larger than 0.7.

\begin{figure}[t]
\centering
\includegraphics[width=8cm]{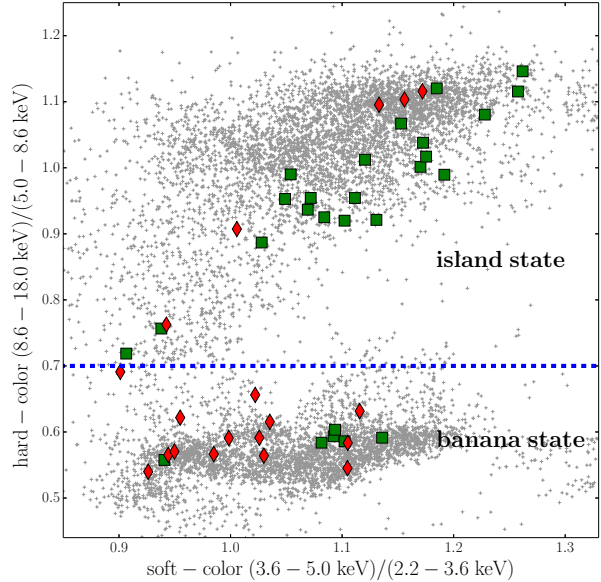}
\caption{Color-Color diagram of 4U 1608--52, in which each point represents 128 s of data.  The blue dashed line represents a hard color equals to 0.7. The non-PRE and PRE bursts are marked with green rectangles and red diamonds, respectively.
}
\label{ccd}
\end{figure}

\section{Results}
First we produced lightcurve for each burst with a time resolution of 1/8 s.
We then performed time-resolved analysis for those time intervals in which the count rate exceeds 25\% of the burst peak.
To improve the statistics, we summed up the time intervals so that the total counts can exceed 35000 photons.
This  means that, in order to have balance in statistics, the bin size has to gradually increase towards the burst tail.
For the spectral fits, we employed two different models: the standard approach \citep[see e.g.][wabs*bbodyrad in XSPEC]{Lewin1993,Kuulkers2003} in which the pre-burst emission is regarded as a background, and the "$f_{\rm a}$" model \citep[wabs*(bbodyrad+$f_{\rm a} \times$persistent flux) in XSPEC]{Worpel2013, intZand2013} in which the "persistent spectrum" is estimated by multiplying the pre-burst spectral shape with a normalization factor "$f_{\rm a}$".
Obviously, the two approaches are equivalent when $f_{\rm a}$ equals to 1.
Since the type-I bursts are located in different spectral states, it is hard to estimate the persistent emissions with an exclusive model.
The models we employed to fit the persistent flux were "wabs*(gauss+bbody+comptt)" and "wabs*(gauss+bbody+diskbb)" \citep{Zhang2013}, and the Gaussian component was set at 6.4 keV \citep{Gierlinski2002}.
It turned out that these two models can describe the persistent data well, and resulted  in an averaged reduced ${\chi}^2$ $\sim$ 1.01 (44dof).

\begin{figure}[t]
\centering
\includegraphics[width=9cm, height=8cm]{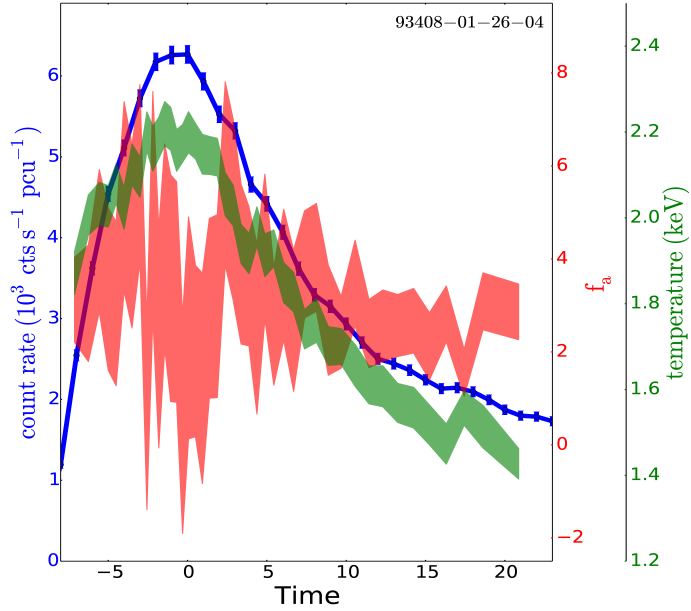}
\caption{An example of a time-resolved spectral analysis with $f_{\rm a}$ model for the burst in observation ID 93408-01-26-04. The blue line shows the count rate vs. the burst evolution. The red and green regions represent the 1 $\sigma$ confidence levels of $f_{\rm a}$ and temperature, respectively.
}
\label{time_resolved}
\end{figure}

An example of the time-resolved analysis is shown in Figure~\ref{time_resolved}, in which the $f_{\rm a}$ model was used.
In Figure~\ref{time_resolved}, the blue line shows the count rate against the burst time, for which we define the time with the maximum count rate as time zero. The red and green regions represent the 1 $\sigma$ confidence interval of $f_{\rm a}$ and temperature, respectively.
Thus, for each burst we can get a series of best-fit parameters along with the evolution of bursts.
Because the purpose of this letter is to distinguish the possible spectral differences between different spectral states, we divided the spectra into the island and banana groups according to the positions of bursts in the CCD (see Figure~\ref{ccd}).
Here we used the reduced ${\chi}^2$ to describe the quality of the fits.
Figure~\ref{standard _chi} shows histograms of the reduced ${\chi}^2$ for different spectral groups when using the standard approach.
Clearly, the two distributions are significantly different and the Kolmogorov--Smirnov test gives a p-value of $1.07\times {10}^{-24}$.
The average reduced ${\chi}^2$ are 1.32 (21 dof) and 1.71 (21 dof) for the spectra in the island and the banana states, respectively.
This result indicates that the fits are in generally poor in the banana state.
In general, once the $f_{\rm a}$ model was introduced, the average reduced ${\chi}^2$ are 1.14 (20 dof) and 1.23 (20 dof) for the island and the banana states, respectively, showing for both with largely improved fits.

Since $f_{\rm a}$ was reported in \citet{intZand2013} to evolve strongly with the burst luminosity,  we studied this trend as well but aimed at uncovering their possible dependence on the spectral states.
Here we estimated the burst luminosity at the surface of the neutron star as $L =(1+z)^2 4\pi \sigma T^{4} N {D}^{2}$, where $\sigma$ is the Stefan--Boltzmann constant, $D$ the distance to the source in the units of 10 kpc, $T$ and $N$ are the color temperature and normalization derived from the spectral fits with the XSPEC model "bbodyrad", respectively. The factor $1+z=(1-2GM/Rc^2)^{-1/2} \sim 1.31$ represents the gravitational redshift correction at the surface of the neutron star.
The result is shown in Figure~\ref{number_L_fa}.
The distribution of $f_{\rm a}$ differs significantly between the island and banana states:  although both have a peak around 2.5, the profile is much broader for the latter.
The average $f_{\rm a}$ for the island and the banana states are $3.2 \pm 0.1$ and $6.8 \pm 0.3$, respectively.
A Kolmogorov--Smirnov test shows that the possibility of having a consistent distribution of $f_{\rm a}$ in the two spectral states is $5.33\times {10}^{-27}$.

To have a clearer trend of the $f_{\rm a}$ evolution against luminosity,  we binned the points in Figure~\ref{number_L_fa} in a bin size of $1.5\times{10}^{37} erg\ s^{-1}$, and within each bin we calculated the weighted average of $f_{\rm a}$.
The result is illustrated in Figure~\ref{L_and_fa}. The red and blue lines represent the weighted average for the bursts in the banana and island states, respectively.
We found that the $f_{\rm a}$ increases gradually with the  burst luminosity in the banana state, which is consistent with the report of \citet{intZand2013}.
However, the bursts in the island state show a very different behavior. When the burst luminosity is smaller than around $1.8\times{10}^{38}\ erg\ s^{-1}$ the $f_{\rm a}$ is similar to that in the banana state, otherwise the $f_{\rm a}$ shows an opposite tendency, i.e. decreases with the increasing burst luminosity.

As shown in Figure~\ref{ccd}, most of bursts in the island state are non-PRE bursts, while in the banana state the PRE bursts are appreciable.
Considering that the photosphere is lifted up when the burst luminosity reaches Eddington limit, it is natural to speculate if the different behaviors as shown in Figure~\ref{L_and_fa} are caused by the photospheric expansion.
To this end we  calculatded the weight averaged $f_{\rm a}$ for both the PRE bursts and non-PRE bursts.
We note that there are no non-PRE bursts having a luminosity larger than $1.8\times {10}^{38} erg\ {s}^{-1}$ in the banana state. Hence, the weight averaged $f_{\rm a}$ in the banana state when the luminosity is larger than this value is actually the contribution from PRE bursts.
In addition, we found that in the island state there are 10 bright bursts (five PRE  and five non-PRE) which have the peak luminosity larger than $1.8\times {10}^{38} erg\ {s}^{-1}$.
We therefore calculated for both types their weight averaged $f_{\rm a}$ and the result is shown in Figure~\ref{L_and_fa} lower panel.
It is obvious that the overall trends are consistent for both types of bursts in the island.

\section{Discussion and summary}
\begin{figure}[t]
\centering
\includegraphics[width=9cm]{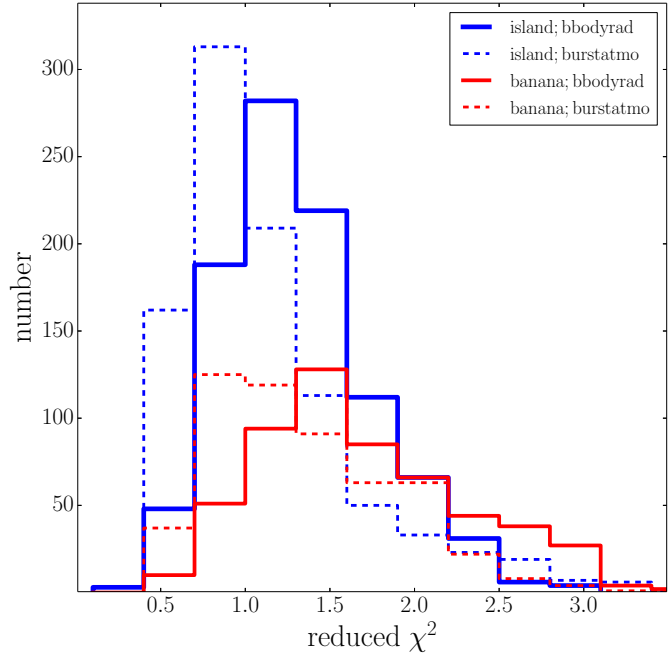}
\caption{The solid (dashed) blue and red histograms show the distributions of the reduced ${\chi}^2$, derived from the wabs*bbodyrad (wabs*burstatmo) model, for the spectra in the island and the banana states, respectively.}
\label{standard _chi}
\end{figure}

\begin{figure}[t]
\centering
\includegraphics[width=9cm]{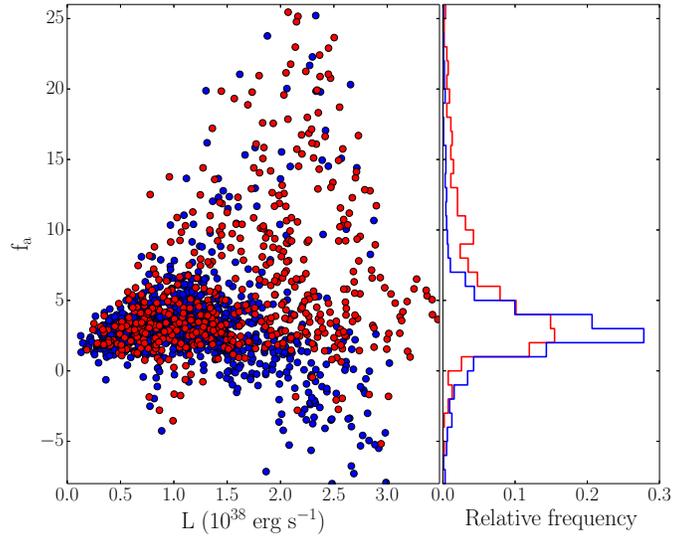}
\caption{left panel: The blue and red points represent the best-fit values of the $f_{\rm a}$ for the bursts in the island state and the banana state, respectively. right panel: The distribution of the points in the left panel. }
\label{number_L_fa}
\end{figure}

In this letter, we investigated the spectral deviations from a blackbody during type-I bursts in different spectral states.
We found that these spectral deviations exist significantly in both of the banana and the island states.
To study whether such deviations are derived from the scattering in the atmosphere of the neutron star as expected by \citet{London1986}, \citet{Madej2004} and \citet{Suleimanov2011, Suleimanov2012}, we fitted the time-resolved spectra using a more accurate model {\rm burstatmo} provided by Suleimanov, during which the persistent flux was regarded as a constant.
The result is shown in Figure~\ref{standard _chi}. Clearly, although the goodness of fit for {\rm burstatmo} model seems to be better than the initial {\rm bbodyrad} model, it is still not sufficient to account for the discrepancy for bursts in different spectral states.

\begin{figure}[t]
\centering
\includegraphics[width=9cm]{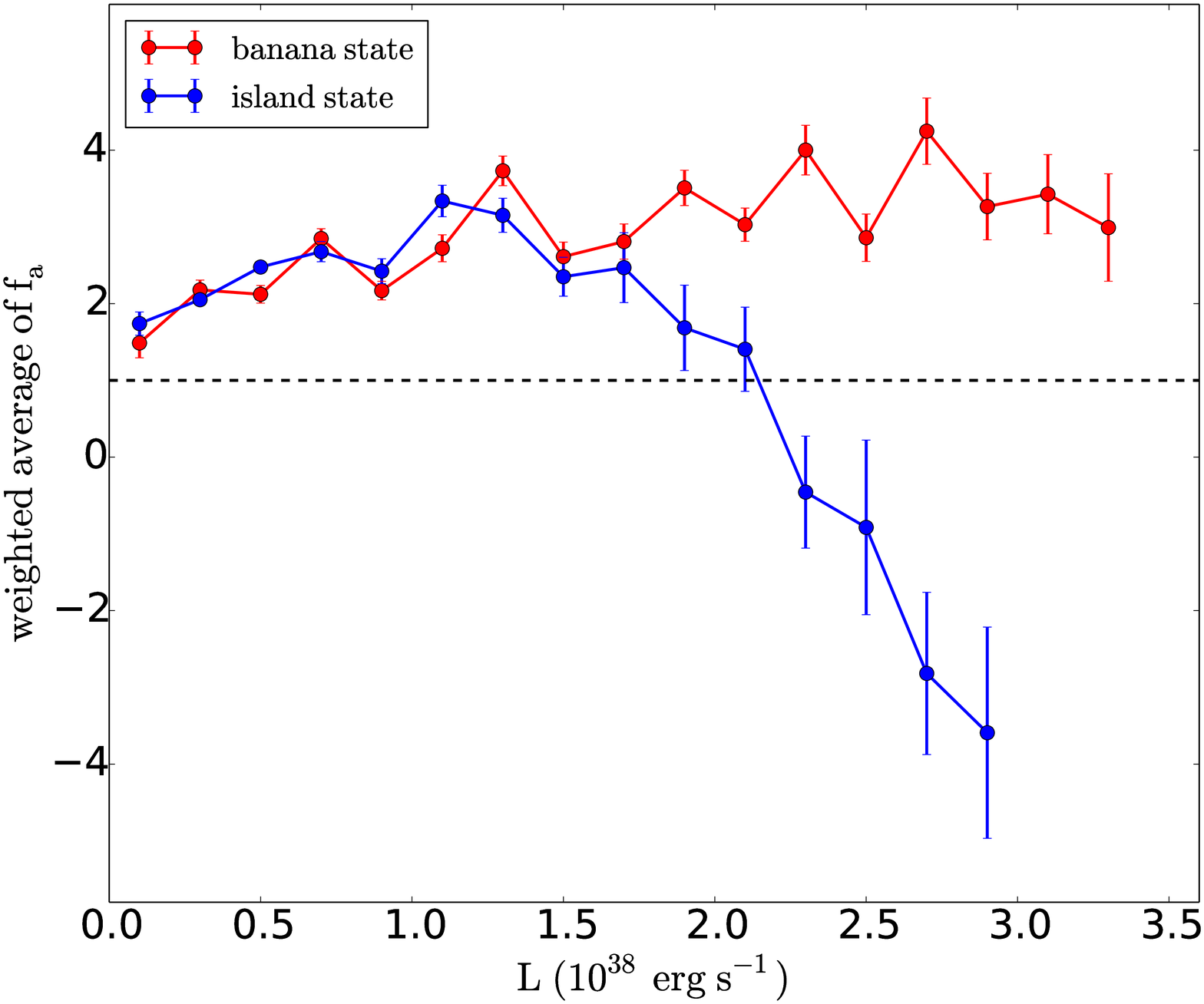}
\includegraphics[width=9cm]{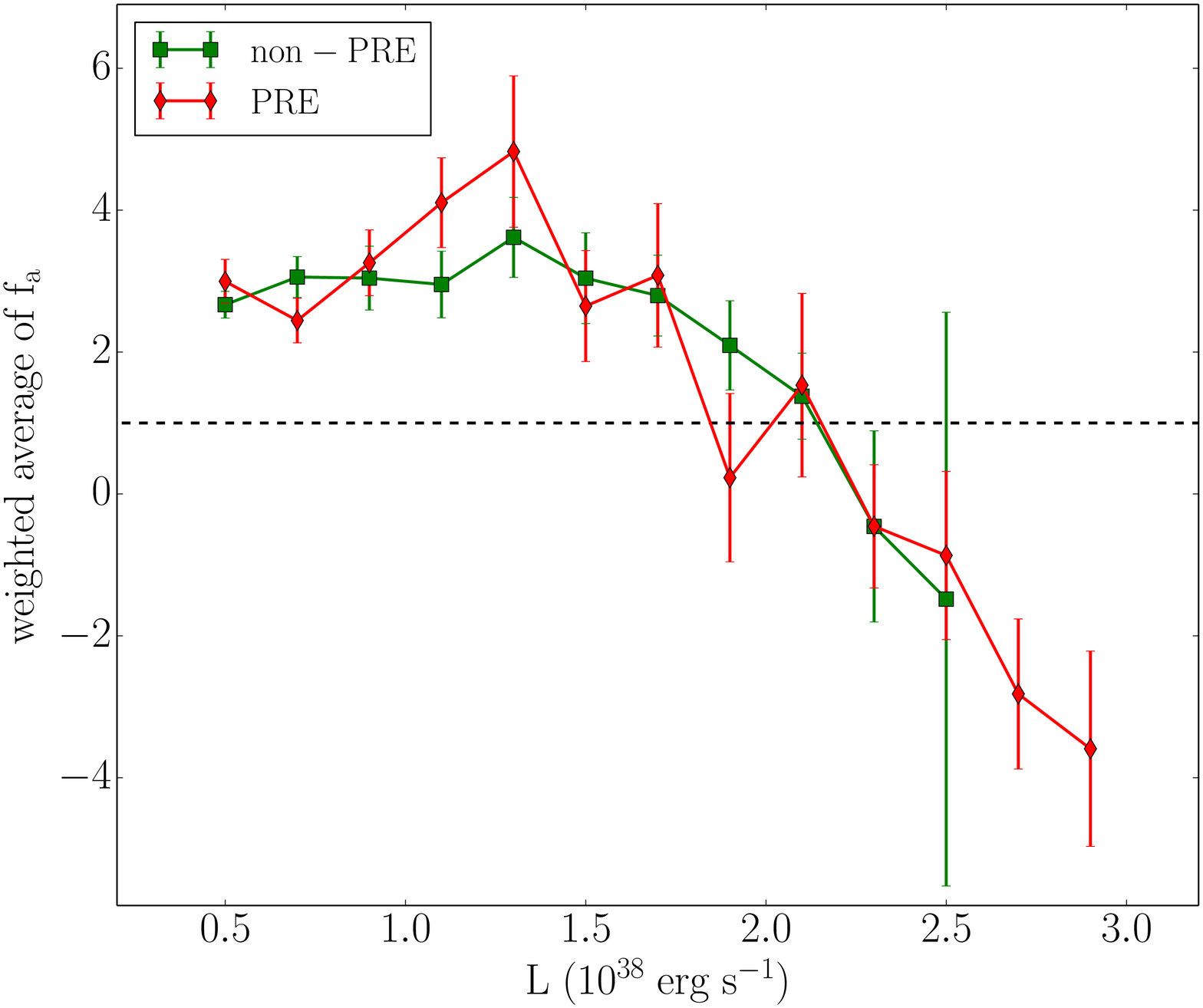}
\caption{Upper panel: Luminosity vs. the weighted average of $f_{\rm a}$ for the spectra in the island and the banana states. The black dashed line shows $f_{\rm a}=1$.
Lower panel: Luminosity vs. the weighted average of $f_{\rm a}$ for PRE bursts and non-PRE bursts in the island state, respectively.}
\label{L_and_fa}
\end{figure}

Technically, a variable persistent flux (i.e. the $f_{\rm a}$ model mentioned above \citep{Worpel2013}) can describe these spectral deviations successfully and improve the fit of burst spectrum.
Figure~\ref{standard _chi} indicates that $f_{\rm a}$ component seems to be more significant in the banana state than what it is in the island state.
Further study has shown that the evolution of $f_{\rm a}$ with increasing burst luminosity varies for bursts in different spectral states.
In the banana state, $f_{\rm a}$ gradually increase with increasing burst luminosity. Instead, in the island state, when the luminosity is larger than  $\sim 1.8\times {10}^{38} erg\ {s}^{-1}$, $f_{\rm a}$ decreases and this tendency is independent of the burst type.
The similar phenomena of $f_{\rm a} >1$ have been reported by \citet{Worpel2013} and \citet{intZand2013}, respectively, and it is explained as an increased persistent flux.
\citet{Worpel2013} attributed it to the increased accretion due to the Poynting-Robertson effect. \citet{intZand2013}
speculated instead that it is due to reprocessing of burst photons by the accretion disc and corona.
Neither of them considered the possibility of the decreased persistent emissions as illustrated in Figure~\ref{L_and_fa}.
We notice that it might be too simple for the assumption in $f_{\rm a}$ model, that the persistent spectral shape keeps unchanged during a burst. For example, by observing the kHz quasi-periodic oscillations, \citet{Peille2014} suggested that some of the $f_{\rm a}$ might be inaccurate for inferring the accretion rate.
Thus, it cannot be ruled out that some fa are artificial (e.g.fa <0 ).
However, if the assumption in $f_{\rm a}$ model does hold, our findings would imply that in the island state the persistent emission is enhanced in the weak bursts, but suppressed in the luminous bursts.
On the contrary, the persistent emission is always enhanced in those bursts born out of the banana state.

In theory, the radiation powered by the type-I bursts will have two important effects on the material orbiting the neutron star. One is to counteract gravity by transferring the radial momentum and the other is to exert a drag on the moving particles (the so-called Poynting-Robertson effect).
Accordingly, the balance of these two opposite effects determines if accretion process will be suppressed or enhanced.
If the latter is more important, more material will fall onto the neutron during bursts, and lead to an increased accretion rate; while if the former is dominant, the material will be blown to infinity and become unbounded, resulting in a suppressed accretion process.


The quantitative calculation has been carried out by \citet{Mishra2013} and \citet{Stahl2013}.
They proved that during type-I bursts a test particle near the innermost stable circular orbit (ISCO) was likely to fall onto the central neutron star.
On the contrary, when the particles are located at the outer part, say, at a distance of around 20 $R_{\rm G}$ or more, they will be hurled out to infinity by a bright burst and be captured by a fainter burst (for details, please see Figure 5 from \citet{Mishra2013}).

Their calculations can naturally explain phenomena that we found, if we assume spectral deviations arise from the interactions between the radiation field of type-I bursts and the accretion disc.
According to the unified model proposed for outbursts in the low mass XRBs, a geometrically thin and optically thick accretion disc (so called "standard disc") extends closer to the compact star in the banana state and is truncated at a larger radius in the island state, during which the corona or jets appear
\citep[see e.g.][and references therein]{Esin1997,Fender2004}.
Although both the Poynting-Robertson effect and the radiation pressure diminish with the distance, the former is more sensitive.
As a result, the minimum luminosity required to eject out the material to infinite, is monotonically decreasing with the increasing distance \citep{Mishra2013,Stahl2013}.
Therefore, the feedback of bursts to the accretion process behaves as a function of the distance between the neutron star and the accretion disc.

In the banana state, since the inner radius of the accretion disc is close to the neutron star, the Poynting-Robertson effect is overwhelming the radial radiation pressure, and leads to an increased accretion rate.
While in the island state the accretion disc is truncated at an outer radius and the escape luminosity is relatively small, the particles in the accretion disc will be ejected out to infinity when bursts are luminous and be dragged onto the neutron star when bursts are less luminous.
Figure~\ref{L_and_fa} lower panel shows that for the non-PRE bursts the $f_{\rm a}$ drops back to 1 at the luminosity $L \sim 2.2 \times {10}^{38} erg\ {s}^{-1}$, which corresponds to around 75\% Eddington luminosity.
Based on the figure 5 from \citet{Mishra2013}, we can deduce that the innermost radius of the accretion disc in the island state is around 16 $R_{\rm G}$, where $R_{\rm G}=GM/{c}^{2}$.

\acknowledgments
We acknowledge support from the Chinese NSFC 11073021, 11133002, 11103020, XTP project XDA 04060604
and the Strategic Priority Research Program "The Emergence of Cosmological Structures" of the Chinese Academy of Sciences, Grant No. XDB09000000.
DFT work is supported by grant AYA2012-39303, and further acknowledges the Chinese Academy of Sciences visiting professorship program 2013-T2J0007.
We thank Suleimanov for providing the {\it Xspec} code for the atmosphere model of neutron stars.

\mbox{}

{}
\end{document}